\documentclass[conference,10pt]{IEEEtran}

\IEEEoverridecommandlockouts

\usepackage{subfigure}
\usepackage{url}
\usepackage{graphicx}
\usepackage{amsmath}
\usepackage{amssymb}

\begin{document}

\title{Determination of Checkpointing Intervals for Malleable Applications\thanks{This work is supported by Department of Science and Technology, India. project ref no. SR/S3/EECE/59/2005/8.6.06}}
\author{
\IEEEauthorblockN{
$^1$K. Raghavendra,
$^2$Sathish S Vadhiyar
}
\IEEEauthorblockA{
$^1$Department of Computer Science and Engineering, Indian Institute of Technology, Madras, India \\
$^2$Department of Computational and Data Sciences, Indian Institute of Science, Bangalore, India \\
raghavendra83@gmail.com, vss@iisc.ac.in
}
}

\maketitle

\begin{abstract}
Selecting optimal intervals of checkpointing an application is
important for minimizing the run time of the application in the
presence of system failures. Most of the existing efforts on
checkpointing interval selection were developed for sequential
applications while few efforts deal with parallel applications where
the applications are executed on the same number of processors for the
entire duration of execution. Some checkpointing systems support
parallel applications where the number of processors on which the
applications execute can be changed during the execution. We refer to
these kinds of parallel applications as {\em malleable}
applications. In this paper, we develop a performance model for
malleable parallel applications that estimates the amount of useful
work performed in unit time (UWT) by a malleable application in the
presence of failures as a function of checkpointing interval. We use
this performance model function with different intervals and select
the interval that maximizes the UWT value. By conducting a large
number of simulations with the traces obtained on real supercomputing
systems, we show that the checkpointing intervals determined by our
model can lead to high efficiency of applications in the presence of
failures.
\end{abstract}

\section{Introduction}
\label{introduction}

With the development of high performance systems with massive number
of processors\cite{top500} and long running scalable scientific
applications that can use the processors for
executions\cite{oliker-petascale-ipdps07}, the mean time between
failures (MTBF) of the processors used for a single application
execution has tremendously
decreased\cite{petrini-systemlevel-ipdps04}. Hence many checkpointing
systems have been developed to enable fault tolerance for application
executions\cite{chen-supportingdynamic-sc06,ruscio-dejavu-ipdps07,fernandez-mobilempi-ppopp06,schulz-implementation-sc04,vadhiyar-srs-ppl2003}.
A checkpointing system periodically saves the state of an application
execution. The application, in the event of a failure, rolls back to
the latest stored or checkpointed state and continues execution.

Recent efforts in checkpointing systems are related to the development
of parallel applications that can change the number of processors
during execution\cite{vadhiyar-srs-ppl2003,fernandez-mobilempi-ppopp06}. We
refer to these kinds of parallel applications as {\em malleable}
applications. Malleable parallel applications are highly useful in
systems with large number of nodes where the resource availability can
vary frequently. In these systems, upon failures of processors used
for processor execution, the application can be made to execute on the
available processors rather than waiting for the failed processors to
be repaired. Malleable applications can also make use of the nodes
that become available during execution.

One of the important parameters in a checkpointing system that
provides fault tolerance is the {\em checkpointing interval} or the
period of checkpointing the application's state. Smaller checkpointing
intervals lead to increased application execution overheads due to
checkpointing while larger checkpointing intervals lead to increased
times for recovery in the event of failures. Hence, optimal
checkpointing intervals that lead to minimum application execution
time in the presence of failures will have to be determined. Large
number of efforts have developed techniques for determining optimal
checkpointing intervals\cite{elnozahy-survey-cmureport96}. These
techniques were primarily developed for sequential applications. They
also consider parallel applications where the number of processors
used by an application remains constant throughout execution. We refer
to these kinds of parallel applications as {\em moldable}
applications.

In this paper, we develop strategies for determining efficient
checkpointing intervals for malleable parallel applications. To our
knowledge, ours is the first effort for malleable applications. Our
work is based on the work by Plank and
Thomason\cite{plank-processorallocation-jpdc01} for finding
checkpointing intervals and suitable number of processors for
executing moldable parallel applications with minimal execution time
in the presence of failures. We extend their Markov models to
incorporate states and transitions that allow reconfiguration of
applications from one processor configuration to another in the event
of failures. The states of our Markov model are automatically
determined from a specified reconfiguration policy. We use different
checkpointing and recovery overheads for different states
corresponding to different number of processors. We also define and
use a new metric for evaluation of a checkpointing interval for
a malleable application, namely, the amount of useful work per unit time
(UWT) performed by the application in the presence of
failures.  Our Markov model is used to estimate the UWT of an
application as a function of checkpointing interval. We use this
performance model function with different checkpointing intervals and
select the interval that maximizes the UWT value. To reduce the
modeling time, we have developed techniques for eliminating low
probable states and transitions, and parallelized the steps for
building the model.

We evaluate the efficiency of our model by using the optimal
checkpointing intervals determined by our model in trace-based
simulations and finding the total amount of useful work performed by
an application in the presence of failures. Our simulations were
conducted with large number of failure traces obtained on both
dedicated batch systems for a 9-year period on 8 parallel systems
and non-dedicated volatile workstations, for three parallel applications,
and with three recovery or rescheduling policies. We show that
checkpointing intervals determined by our models lead to greater than
80\% application efficiency in terms of
useful work
performed by the application in the presence of failures.

Following are our primary contributions. \\
1. Developing a model for execution of malleable applications
   based on the the model by Plank and Thomason for moldable
   applications. This includes significant extensions to the original
   model including different definitions for the states of the model,
   and automatic determination of the states and transition
   probabilities based on reconfiguration or rescheduling policies. \\
2. Definition of a new metric for evaluating checkpointing
  intervals for malleable applications. \\
3. Optimizing the model by elimination of states and employing
  parallelism. \\
4. Extensive simulations with real world failure traces for
  different applications and with different rescheduling policies.

Section \ref{background} summarizes the work by Plank and Thomason on
modeling moldable applications. Section \ref{malleable} describes in
detail our model for malleable parallel applications and the metric
used for evaluation of checkpointing intervals determined by our
model. Section \ref{impop} explains the optimizations of the model
framework. Section \ref{rescheduling} describes the rescheduling
policies used by our model. In Section \ref{experiments}, we describe
the various simulation experiments we conducted to show the efficiency
of the checkpointing intervals determined by our models. Section
\ref{related} presents related work in the area. Section
\ref{conclusions} gives conclusions and Section \ref{futurework}
presents future work.

\section{Background: Checkpoint Intervals for Moldable Applications}
\label{background}

The work by Plank and Thomason\cite{plank-processorallocation-jpdc01}
developed a finite-state Markov chain based performance model to
characterize the execution of long running moldable parallel
application in the presence of failures. They use the model to find
the checkpointing interval, $I$, and the number of processors, $a$,
for execution of a long running application in a system with $N$
($N>a$) processors. Their goal is to minimize the running time of the
application in the presence of failures. {\em Spare processors} are
defined as candidate processors for replacing a processor that failed
during application execution. The number of spare processors, $S$, is
$N-a$. $L$ is the checkpointing latency or the total time spent for
checkpointing and $C$ is defined as the checkpointing overhead or the
extra time incurred by the application due to
checkpointing. Typically, $C<L$ due to optimizations performed in
checkpointing systems. $R$ is the time spent in recovery from a
failure. The model assumes exponential distribution for
inter-occurrence times of failures and repairs for a
processor. $\lambda$ denotes the failure rate and $\theta$ denotes the
repair rate for a single processor. Given a trace of failures and
repairs of a processor, the mean time to failure (MTTF) of the
processor is calculated as the average of times between failures of
the processor. The mean time to repair (MTTR) of a processor is
calculated as the average of times from when the processor fails to
when the processor is available for execution. For a multi-processor
system, $\lambda$ and $\theta$ are calculated as the reciprocal of the
average of MTTFs and MTTRs, respectively, for all processors.

\begin{figure}
\centering
\includegraphics[width=0.45\textwidth]{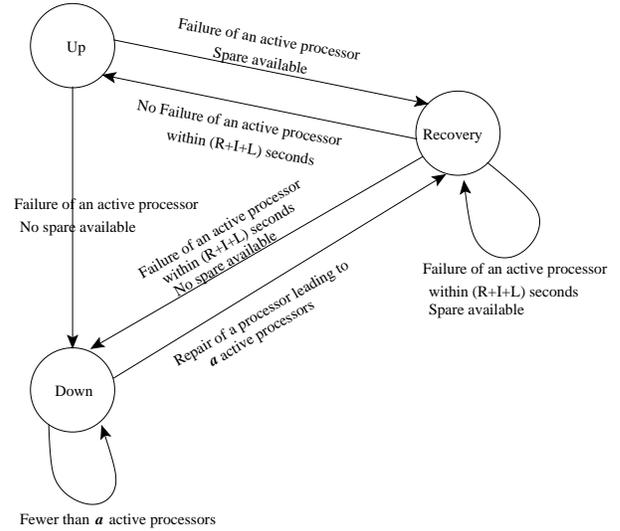}
\caption{States and Transitions for Markov model for moldable
  applications, $M^{mold}$}
\label{markov-mold}
\end{figure}

The Markov chain, $M^{mold}$, consists of three types of states,
namely, {\em up}, {\em down} and {\em recovery}, as shown in Figure
\ref{markov-mold}. The application is in
an {\em up} state if at least $a$ processors are available for
execution. If one of the processors used for the execution of the
application fails and the total number of functional processors
remaining in the system is less than $a$, the application is halted
and is considered to be in a {\em down} state. The application remains in
this state until some of the failed processors are repaired and at
least $a$ processors become available for execution again. In this
case, the application goes to a {\em recovery} state. The application also
enters a {\em recovery} state from an {\em up} state if after the failure of a
processor used during the execution, the number of remaining
functional processors is at least $a$. In the {\em recovery} state, the
application tries to recover from the previous checkpoint, spending
$R$ seconds for rollback to the checkpointed state, and remains in the
{\em recovery} state until it creates a new checkpoint after $(I+L)$
seconds. If during the $(R+I+L)$ seconds, none of the processors fail, the
application enters an {\em up} state and continues execution. If one of the
processors fails during the recovery and spares are available for
replacement of the failed processor, the application enters another
{\em recovery} state and restarts the recovery process. However, if spares
are not available, the application enters a {\em down} state.

The Markov model, $M^{mold}$ consists of $S+1$ {\em up} states, $S$
{\em recovery} states and $a$ {\em down} states. An {\em up} state
denoted by $[U:s]$, $0\leq s\leq S$, corresponds to application
execution on $a$ processors with $s$ spare processors in the system at
the time the state is entered. A {\em recovery} state denoted by
$[R:s]$, $0\leq s< S$, corresponds to application recovery on $a$
processors with $s$ spares available at the start of the
recovery. When the application exits an {\em up} state due to failure
of one of the $a$ processors with $s+1$ spares available, it goes to
$[R:s]$ state after replacing the failed processor with one of the
$s+1$ spares. After a span of $(R+I+L)$ seconds in a {\em recovery}
state, $[R:s1]$, the application enters an {\em up} state, $[U:s2]$,
where $s2$ is the number of spares in the system when the {\em up}
state is entered. If an application exits an {\em up} state due to
failure of one of the $a$ processors and the total number of
functional processors in the systems is $a-1$, the {\em down} state,
$[D:a-1]$, is entered. The {\em down} state denoted by $[D:p]$, $0\leq
p<a$, represents the system with only $p$ processors available. The
{\em recovery} state, $[R:0]$, is entered from a {\em down} state,
$[D:a-1]$, after repair of a failed processor resulting in exactly $a$
functional processors.

\begin{figure}
\centering
\includegraphics[width=0.45\textwidth]{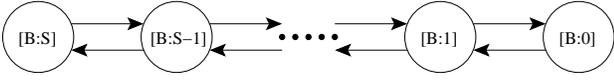}
\caption{Birth-Death Markov Chain $S^{\tau}$}
\label{bd-markov}
\end{figure}

The probabilities of transitions from the states in $M^{mold}$ are based
on the number of functional spares available after the exits of the
states. These probabilities are calculated using a  birth-death Markov
chain,  $S^{\tau}$, that helps find the probability of starting
with $i$ spares and ending with $j$ spares, $0\leq i,j\leq S$, after
$\tau$ seconds. The Markov chain, $S^{\tau}$, consists of $S+1$
states, denoted $[B:s]$, $0\leq s\leq S$,
, as shown in Figure
\ref{bd-markov}.
Each state, $[B:s]$, corresponds to $s$ functional
spares and $S-s$ processors under repair. Transition out of a state
$[B:s]$ is either to the state $[B:s-1]$ corresponding to failure of a
single processor with probability $s\lambda$ or to $[B:s+1]$ corresponding
to repair of a processor with probability $(S-s)\theta$. The states
are numbered $1$ to $S+1$ from left to right such that state $i$
represents $S-i+1$ functional spares.

A $(S+1)\times (S+1)$ square matrix, {\bf R}, of instantaneous
probabilities is defined as:
\begin{equation}
{\bf R} = 
\begin{bmatrix}
-S\lambda & S\lambda & \cdots & 0 & 0 \\
\theta & -((S-1)\lambda + \theta) & \cdots & 0 & 0 \\
\vdots & \vdots & \vdots & \vdots & \vdots \\
0 & 0 & \cdots & -(\lambda + (S-1)\theta ) & \lambda \\
0 & 0 & \cdots & -S\theta & S\theta \\
\end{bmatrix}
\label{Rmatrix}
\end{equation}

The matrix {\bf R} is used to calculate a $(S+1)\times (S+1)$ matrix,
${\bf Q}^{S,\tau} = [q_{i,j}^{S,\tau}]$, shown in Equation
\ref{q-tau}, where the $[i,j]$ entry is the probability
that the Markov chain $S^{\tau}$ starting in state $i$ enters state
$j$ after $\tau$ seconds. Thus $q_{S-i+1,S-j+1}^{S,\tau}$ is the
probability of starting with $i$ functional spares and ending with $j$
spares after $\tau$ seconds.
\begin{equation}
{\bf Q}^{S,\tau} = expm({\bf R}\tau )
\label{q-tau}
\end{equation}
$expm({\bf R}\tau )$ is the matrix exponential of ${\bf R}\tau$.

If $f_{\tau}$ is the probability density function of the TTF (time to
failure) random variable $\tau$, the likelihoods of transitions
between the states in $S^{\tau}$ are given by the $(S+1)\times (S+1)$
matrix:
\begin{equation} 
[q_{ij}^S] = \int_{t}{\bf Q}^{S,t}f_{\tau}(t)dt
\label{integral}
\end{equation} 

The transition probabilities in the original Markov model, $M^{mold}$,
are calculated using the ${\bf Q}^{S,\tau}$ and $[q_{ij}^S]$ matrices
of the birth-death Markov chain, $S^{\tau}$. We illustrate the
calculations for probabilities of transitions from the {\em recovery}
states. For successful transitions from {\em recovery} to {\em up}
states in $M^{mold}$, a failure must not occur within $\delta = R+I+L$
seconds. The probability of no active processor failure during the
interval $[0,\delta]$ is $e^{-a\lambda \delta}$. The probabilities of
the specific {\em up} states after transitions are given by ${\bf
  Q}^{S,\delta}$, obtained by substituting $\delta$ for $\tau$ in
Equation \ref{q-tau}. Thus, the probability of transition from $[R:i]$
to $[U:j]$ is $(e^{-a\lambda
  \delta})(q_{S-i+1,S-j+1}^{S,\delta})$. The probability of an active
processor failure within $\delta$ seconds is $1-e^{-a\lambda
  \delta}$. The failure results in a transition to a {\em recovery}
state or the {\em down} state $[D:a-1]$ depending on the number of
spares. The probability density function of the TTF random variable
$\tau$ is $a\lambda^{-a\lambda \tau}$ and is conditioned on $\tau$
being in the interval $[0,\delta]$. Thus substituting $f_{\tau}(t) =
\frac{a\lambda^{-a\lambda t}}{1-e^{-a\lambda \delta}}$ in Equation
\ref{integral} and integrating over the interval $[0,\delta]$, the
matrix of likelihoods, ${\bf Q}^{Rec,S}=[q_{ij}^{Rec,S}]$, of
transitions between the
states in $S^{\tau}$ is calculated. The probability of a transition
from $[R:i]$ to $[R:j]$ in $M^{mold}$ is then $(1-e^{-a\lambda
  \delta})q_{S-i+1,S-j+1}^{Rec,S}$ and to $[D:a-1]$ is
$(1-e^{-a\lambda \delta})q_{S-i+1,S+1}^{Rec,S}$. 
Similar calculations are used to find the transition probabilities
from the {\em up} and {\em down} states and are explained in
\cite{plank-processorallocation-jpdc01}. For finding the transition
probabilities from the {\em up} states, the matrix of likelihoods,
${\bf Q}^{Up,S}=[q_{i,j}^{Up,S}]$, is used. ${\bf Q}^{Up,S}$ is calculated by
substituting $f_{\tau}(t) = a\lambda e^{-a\lambda t}$ in Equation
\ref{integral} and integrating over the interval $[0,\infty]$.
The integral equations for
the calculations of transition probabilities from the {\em up} and
{\em recovery} states, using Equation \ref{integral}, are solved by
computing the
eigen values and eigen vectors of the ${\bf R}$ matrix, shown in
Equation \ref{Rmatrix}. These solutions are also described in detail
in \cite{plank-processorallocation-jpdc01}. The calculated
probabilities of transitions in $M^{mold}$ are represented by a square
matrix, ${\bf P}^{mold}$, with the number of rows or columns equal to
the total number of {\em up}, {\em recovery} and {\em down} states,
and row of ${\bf P}^{mold}$ corresponds to a state of $M^{mold}$ such
that $P_{i,j}^{mold}$ is the probability of transition from state $i$
to state $j$ in $M^{mold}$.

Each transition, $i,j$, in $M^{mold}$ is also weighted by $U_{i,j}$,
the average amount of useful time or uptime spent by the application
in the state corresponding to the start of the transition and
$D_{i,j}$, the non-useful or down time spent by the application in the
state. The uptime is the time spent by the application performing
useful work and is equal to the failure-free running time of the
application not enabled with checkpointing. The down time includes the
time spent in checkpointing, $C$, recovery, $R$, recomputation of work
lost due to a failure, and the time spent in the {\em down}
states.
For example, for a transition from a {\em recovery} state $[R:i]$ to
an {\em up} state $[U:j]$, the useful time, $U_{i,j} = I$ and the down time,
$D_{i,j}=R+L$.
For a transition to another
{\em recovery} state, $[R:j]$, a failure must have occurred within the
$\delta$ seconds. Thus, the useful time, $U_{i,j} = 0$ and the down
time, $D_{i,j}= \frac{1}{a\lambda}-\delta \frac{e^{-a\lambda
    \delta}}{1-e^{-a\lambda \delta}}$, the MTTF (mean time to failure)
conditioned on failure within $\delta$ seconds.
The useful and down
times for the other transitions are calculated similarly and are shown
in the work  by Plank and
Thomason\cite{plank-processorallocation-jpdc01}. Thus square matrices,
${\bf U}$ and ${\bf D}$ are constructed corresponding to the
transition matrix, ${\bf P}^{mold}$.

The long-run properties of $M^{mold}$, where $M^{mold}$ is taken
through a large number of transitions, are used to find the long-run
probability of the occupancy of state $i$. This is given by the entry
$\pi_i$  in the unique solution of the matrix equation, $\Pi = \Pi
{\bf P}^{mold}$. If $M^{mold}$ is taken through $n$ transitions
randomized according to the transition probabilities, and if $n_i$ is
the number of occurrences of state $i$ during those transitions, then
\begin{equation}
\pi_i = \lim_{n\to \infty} \frac{n_i}{n+1}
\end{equation}
Since each visit to state $i$ is followed by probabilistic selection
of an exit transition, the limiting relative frequency of occurrence of
transition $i\rightarrow j$ is the joint probability
$\pi_iP^{mold}_{i,j}$. Thus, for a long-running task,  $U_{i,j}\pi_i
P^{mold}_{i,j}$ and $D_{i,j}\pi_i P^{mold}_{i,j}$ are the expected
contributions of useful and non-useful times, respectively, due to the
relative frequency of transition $i\rightarrow j$. The availability,
$A_{a,I}$, for a given number of processors, $a$, used for execution
and a given checkpointing interval, $I$, is the ratio of the mean
useful time spent per transition to the mean total time per transition
and is calculated as:
\begin{equation}
A_{a,I} =
\frac{\sum_{i,j}U_{i,j}\pi_i P^{mold}_{i,j}}{\sum_{i,j}(U_{i,j}+D_{i,j})\pi_i
  P^{mold}_{i,j}}
\end{equation}
By trying different values for a and I, the work by Plank and Thomason
chooses $a$ and $I$ that minimize the expected execution time of the
application  in the presence of failures, $RT_a/A_{a,I}$, where $RT_a$
is the estimated failure-free execution time.

\section{Checkpoint Intervals for Malleable Applications}
\label{malleable}

Few checkpointing systems enable malleable applications where the
number of processors used for execution can be changed during the
execution. In our model for executing malleable applications in a
system consisting of $N$ processors, instead of choosing a fixed number
of processors, $a$, at the beginning of execution, the number of
processors for execution is chosen at different points in application
execution. The number of processors chosen at a particular point of
execution is a function of the number of functional processors
available at that point and is specified by a {\em rescheduling
  policy}. The rescheduling policy is denoted by a vector, $rp$, of
size $N$ where $rp_i$ denotes the number of processors that will be
selected for application execution given $i$ functional
processors. The vector $rp$ is specified as input to our
model. Section \ref{rescheduling} explains the different kinds of rescheduling
policies employed in this work.

In this section, we describe our Markov model for execution of
malleable applications, the inputs and outputs of our model, and the
process of selecting the best checkpointing intervals for malleable
applications. 

\subsection{Markov Model}
\label{markovformalleable}

\begin{figure}
\centering
\includegraphics[width=0.4\textwidth]{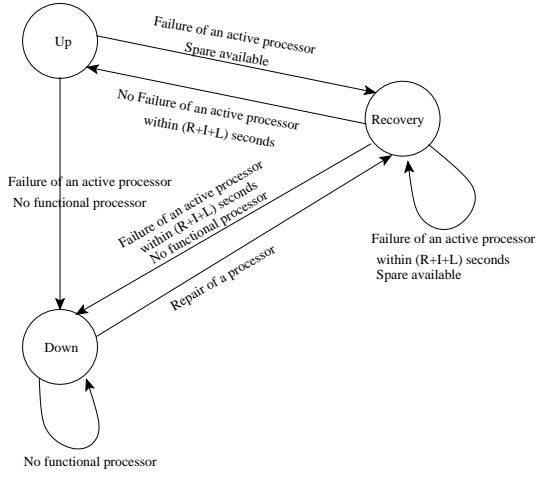}
\caption{States and Transitions for Markov model for malleable
  applications, $M^{mall}$}
\label{markov-mall}
\end{figure}

Our model for malleable applications also involves three kinds of
states, namely, {\em up}, {\em down} and {\em recovery}. In our model,
a long-running
malleable parallel application initially starts execution on $rp_i$
number of processors with $i$ total number of functional processors in
the system at the beginning of execution. At this point, the
application is considered to be in an {\em up} state. The application, after
every $I$ seconds, stores a checkpoint, incurring an application
overhead of $C_{rp_i}$ corresponding to $rp_i$ processors. For our
work, we assume that the checkpoint overhead, $C_{rp_i}$ is equal to
the latency, $L_{rp_i}$. When a processor used by the executing
application fails, the application is recovered on $rp_j$ processors
corresponding to $j$ total number of functional processors available
at the time of failure. Thus the application makes a transition to a
{\em recovery} state. Recovery involves redistribution of data in the
application from the previous processor configuration to the new
configuration. Unlike for moldable applications, the time taken for
recovery, $R_{k,l}$, depends on the number of processors, $k$, used by
the application before failure and the number of processors, $l$, on
which the application will be recovered. In the {\em recovery} state, the
application tries to recover from the previous checkpoint and create a
new checkpoint after $(R_{rp_i,rp_j}+I+C_{rp_j})$ seconds. If during
this time, none of the processors involved in the recovery fails, the
application enters an {\em up} state. If one of the processors involved in
recovery fails, the application restarts the recovery process in
another {\em recovery} state. Thus in our model, the checkpointing overhead
and the rescheduling cost vary for different states and transitions
corresponding to the number of active processors used for execution
and recovery at different points of execution. The application goes to
a {\em down} state if the total number of functional processors in the
system is less than the minimum number of processors required for
execution. Without loss of generality, for this work, we assume that
the application can execute on a single processor. Hence there is only
one {\em down} state in our model corresponding to failure of all processors
in the system.
The states and transitions for our model for malleable
applications are illustrated in Figure \ref{markov-mall}. Comparing
with Figure \ref{markov-mold}, the corresponding figure for moldable
applications, we find that the primary difference is in terms of the
transitions to and from the {\em down} state.

In our Markov model, $M^{mall}$, for malleable applications, an {\em up}
state is denoted by $[U:a,s]$ where $1\leq a\leq N$ is the number of
active processors used for execution of application in the state and
$0\leq s \leq S$ is the number of spare processors corresponding to
the state. $S$ ($=N-a$) is the maximum number of spares in the system
corresponding to $a$ active processors. Thus the total number of {\em up}
states in $M^{mall}$ is equal to $\frac{N(N+1)}{2}$. A {\em recovery} state
is similarly denoted by $[R:a,s]$ where $1\leq a\leq N$ is the number
of active processors on which the application is recovered and $s$ is
the number of spare processors corresponding to the state. The
{\em recovery} state $[R:a,s]$ corresponds to a unique element in the
rescheduling policy vector, $rp$. Specifically, $[R:a,s]$ corresponds
to $(a+s)^{th}$ element in $rp$ where $(a+s)$ denotes the total number
of functional processors in the system and $rp_{(a+s)}(=a)$ denotes the
number of processors selected for recovery or execution. Since the
size of $rp$ vector is $N$, the total number of {\em recovery} states in
$M^{mall}$ is $N$. The exact {\em recovery} states in our model are thus
dependent on the specified rescheduling policy and are dynamically
determined.\footnote{Note that the number of {\em up} states is not related to
  the entries in the rescheduling policy vector, $rp$. The number of
  {\em up} states is not equal to $N$, the number of {\em recovery} states, since
  after recovery on $a_1$ processors with $s_1$ spares as dictated by
  the $rp$ vector, the application can enter an {\em up} state $[U:a_1:s_2]$
  with different number of spares, $s_2$, available at the start of
  the {\em up} state.}

Relating this Markov model, $M^{mall}$, to the Markov model for
moldable applications, $M^{mold}$, the {\em up} states in $M^{mall}$ contain
the {\em up} states in $M^{mold}$ that correspond to given number of active
processors, $a$, for all possible values of $a$. In $M^{mold}$, the
{\em recovery} states for application recovery on $a$
processors correspond to the number of spares available at the time of
recovery. In $M^{mall}$, the {\em recovery} states correspond to the total
number of functional processors available at the time of
recovery and the actual number of processors used
for recovery can vary in different states.

The probabilities of transitions between the states in $M^{mall}$ are
represented by a square matrix, ${\bf P}^{mall}$, with the number of
rows or columns equal to the total number of {\em up}, {\em recovery} and
{\em down} states. In order to fill the entries in the matrix for
transitions from the states corresponding to application execution or
recovery on specific number of processors, $a$, with $S$ ($=N-a$)
number of spares, a birth-death Markov chain, $S^{\tau}$, and the
corresponding ${\bf Q}^{S,\tau}$, ${\bf Q}^{Up,S}$ and ${\bf
  Q}^{Rec,S}$ matrices are constructed in the same way as in the
model, $M^{mold}$,  for moldable applications. However, unlike for
$M^{mold}$, where a single birth-death Markov model was constructed
for modeling execution on a fixed number of active processors, $a$
(with $S$ spares), we construct $N$ such birth-death Markov models for
$M^{mall}$ corresponding to execution on $N$ possible number of active
processors with the corresponding number of spares and obtain $N$
corresponding ${\bf Q}^{s,\tau}$, ${\bf Q}^{Up,s}$ and ${\bf
  Q}^{Rec,s}$ matrices, where $0\leq s<N$.

These probabilities of transitions starting with a certain number of
spares and ending with another number of spares are used to calculate
the entries of the ${\bf P}^{mall}$ matrix. For $a$ number of active
processors used for execution or recovery, and $S(=N-a)$ spares, an
$[i,j]$ entry in the matrix ${\bf P}^{mall}$ corresponding to a {\em
  recovery}-to-{\em up} transition is calculated based on the
$[S-i+1,S-j+1]$ entry in the ${\bf Q}^{Rec,S}$ similar to the
calculation of $[i,j]$ entries in the ${\bf P}^{mold}$ matrix for
moldable applications. However, unlike in the construction of ${\bf
  P}^{mold}$, an entry $[i,j]$ in the ${\bf P}^{mall}$ corresponding
to a transition to a {\em recovery} state cannot be calculated
directly from the corresponding $[S-i+1,S-j+1]$ entry in the ${\bf
  Q}^{Up,S}$ or ${\bf Q}^{Rec,S}$ matrices. This is because the ending
state of the transition to a {\em recovery} state in $M^{mall}$ not
only depends on the number of spares, but also on the number of active
processors used for recovery. The number of active processors in turn
depends on the rescheduling policy given by the rescheduling policy
vector, $rp$. For example, for a transition from an {\em up} state,
$[U:a1:s1]$ to a {\em recovery} state with $s2$ spares, the
$[S-s1+1,S-s2]$ entry in the ${\bf Q}^{Up,S}$ matrix is used for the
calculation of probability of transition to the {\em recovery} state
$[R:rp_{(s2+a1-1)},s2]$ in the ${\bf P}^{mall}$ matrix. $(s2+a1-1)$ is
the total number of available functional processors at the start of
the recovery corresponding to $s2$ spares. This is the sum of the
number of spare processors and the number of remaining active
processors used for execution in the {\em up} state. The number of
remaining active processors at the end of the {\em up} state is $a1-1$
since one of the active processors at the beginning of the {\em up}
state has failed during the execution causing the application to
transition to the {\em recovery} state. $rp_{s2+a1-1}(\le (s2+a1-1))$
is the number of processors on which the application will be recovered
corresponding to the total number of functional processors,
$(s2+a1-1)$, and is specified in the rescheduling policy.

\subsection{Useful Work per Unit Time}
\label{uwt}

For a moldable parallel application, the best checkpointing interval,
$I$, for a given number of processors, $a$, is selected by trying
different values for $I$, obtaining availability, $A_{a,I}$, for each
value using the Markov model, and choosing the interval for which
$RT_a/A_{a,I}$ is minimum. Here, $RT_a$ is the estimated failure-free
execution time, and $RT_a/A_{a,I}$ is the estimated executed time in
the presence of failures for the application. However this approach
cannot be used for finding the best checkpointing interval for
malleable parallel applications. This is because the number of
processors used for execution changes during the execution and hence
changes in the various states of our model, $M^{mall}$. Thus a single
failure free running time corresponding to a certain number of
processors cannot be used.

For malleable applications, we use a metric called total useful
work per unit time (UWT) defined as:
\begin{equation}
UWT = \frac{W_I}{U_I+D_I}
\end{equation}
where $W_I$ is the total amount of useful work, and $U_I$ and $D_I$
are the total up and down times for a checkpointing interval, $I$. The
up and down times are calculated as described in Section
\ref{background} for moldable applications. For a state visited in our
model, $M^{mall}$, corresponding to certain number of processors, $a$,
let $uptime$ be the total up time spent in the state. The amount of
useful work performed in the state, $work$,  is the estimated amount
of computations that can be performed on $a$ processors in $uptime$
seconds spent in the state and is calculated as
$work=workinunittime_a\times uptime$ where $workinunittime_a$ is the
amount of computations that can be performed on $a$ processors in one
second. For example, for iterative regular parallel applications,
$workinunittime_a$ can be the number of iterations that can be
completed by the application in one second on $a$ processors. The
vector $workinunittime$ for different number of processors is given as
an input to our model, $M^{mall}$. $W_I$ for the complete model for a
specified checkpointing interval, $I$, is calculated by accumulating
the amount of useful work performed in all the states visited in the
model during execution.

Thus, a transition, $(i\to j)$, in $M^{mall}$ is weighted by the
average up time, $U_{i,j}$, down time, $D_{i,j}$, and the amount of
useful work performed by the application, $W_{i,j}$, in the state
corresponding to the start of the transition. The square matrices,
${\bf U}$, ${\bf D}$ and ${\bf W}$ are constructed corresponding to
the transition matrix, ${\bf P}^{mall}$. Using the long-run properties
of $M^{mall}$, and calculating $\pi_i$ as in $M^{mold}$, the amount of
useful work per unit time, $UWT_I$, for a given checkpointing
interval, $I$, is calculated as:
\begin{equation}
UWT_I =
\frac{\sum_{i,j}W_{i,j}\pi_i P^{mall}_{i,j}}{\sum_{i,j}(U_{i,j}+D_{i,j})\pi_i
  P^{mall}_{i,j}}
\end{equation}

\subsection{Selecting Checkpointing Intervals}
\label{selectingI}

The user specifies the following parameters for building our model,
$M^{mall}$: \\
1. $N$, $\lambda$ and $\theta$ corresponding to the system, \\
2. a vector $C$ corresponding to checkpointing of the
application for different number of processors, \\
3. a matrix $R$ corresponding to recovery from a certain number of processors to a
different number of processors, \\
4. a vector $workinunittime$ for the application, \\
5. a vector $rp$ specifying the rescheduling policy, and \\
6. a checkpointing interval, $I$. \\
The model is used to obtain $UWT_I$ for a checkpointing interval,
$I$. By trying different values for $I$, the user chooses the interval
that maximizes the expected useful work performed by the application
per unit time. 

Most of the parameters necessary for the user to select efficient
checkpointing intervals can be easily derived. For $N$, the user
specifies the total number of processors available in the
system. Given a failure trace for a system, $\lambda$ and $\theta$ can
be derived by observing the times between any two consecutive failures
and the times taken for repairs of a failed system, respectively, and
calculating the averages of the times. We have developed programs that
can be used with standard failure traces to automatically calculate
$\lambda$ and $\theta$.

The vectors, $workinunittime$ and $C$, and the matrix, $R$, are
obtained by benchmarking the applications. Our work on checkpointing
intervals is primarily intended for long-running large scientific
applications. Such applications are typically benchmarked by the users
for different problem sizes and number of processors for application
development and performance improvement. The user links his
application with a checkpointing library, executes parts of the application for
different configurations, and collects the times taken for the
executions. For example, for an iterative application, the user executes
the application for few iterations, finds the time taken for execution
of the iterations, and obtains the number of iterations executed in a
second or work performed by the application in unit time.
For checkpointing and recovery overheads, the user
obtains the times by inserting time stamps at the beginning and end of
the checkpointing and recovery codes, respectively, in the
checkpointing library. The checkpointing and recovery codes are
invoked as functions that are inserted in the application codes in
many checkpointing
systems\cite{fernandez-mobilempi-ppopp06,vadhiyar-srs-ppl2003}, and
hence can be easily identified by the user. For obtaining recovery
overheads, the user can induce failures to an executing application on
a certain number of processors and continue on a different number of
processors. After obtaining the work performed in unit time,
checkpointing and recovery overheads, for a certain set of processor
configurations, the user can construct the vectors, $workinunittime$
and $C$, and the matrix, $R$, respectively, for all number of
processors using either simple techniques including average, maximum
or minimum or complex strategies like extrapolations. The vectors,
$workinunittime$ and $C$, and the matrix, $R$, are constructed only
once for a given application and system and are used for multiple
executions.

The complexity of constructing the rescheduling policy vector, $rp$,
depends on the complexity of the rescheduling policy that the user
wants to implement. A simple rescheduling policy can be to continue
the application on all the available number of processors. In this
case, the rescheduling policy vector will simply contain integers ranging
from $1$ to $N$, the total number of processors in the system. Some
rescheduling policies are discussed in Section \ref{rescheduling}.

\section{Implementation and Optimizations}
\label{impop}

We have developed MATLAB scripts for implementing the process of
selecting checkpointing intervals for malleable applications.
Our scripts are based on the MATLAB scripts
developed for moldable applications by Plank and
Thomason\cite{plank-processorallocation-jpdc01}.

The number of {\em up} states in our model, $M^{mall}$, is $O(N^2)$. In order
to reduce the number of {\em up} states and hence the space complexity
and execution time of our model, we eliminate an {\em up} state if the
probabilities of transitions to the {\em up} state is less than a threshold,
$thres$. Large values of $thres$ will result in
elimination of many {\em up} states and will result in high modeling
errors. Small values of $thres$  will not eliminate significant
number of states and hence cannot significantly reduce the space and
time complexities of the model. Hence we choose a value for
$thres$ that results in small modeling errors due to elimination of
states and significant number of eliminated states.
We conducted 750 different experiments by building our model with different
failure traces corresponding to different $\lambda$s, different
checkpointing intervals, $I$, and different application parameters, $R$ and
$C$. For each of these experiments, we used eight different thresholds for
$thres$, and executed the resulting reduced models. We computed a
score for a threshold for an experiment as:
\begin{equation}
score = \alpha (1.0-threserror)+\beta (elims)
\end{equation}
where $threserror$ (between $0.0-1.0$) is the model error due to
elimination of the {\em up} states and is calculated as the percentage
difference between the $UWT$ of the original model, $M^{mall}$, and
the $UWT$ of the reduced model with some {\em up} states eliminated. $elims$
is the number of eliminated {\em up} states. $\alpha$ is the weight
associated with the modeling error and $\beta$ is the weight
associated with the number of eliminated {\em up} states. Large values of
$\alpha$ result in high scores for $thres$ values that yield small
modeling errors while large values of $\beta$ result in high scores
for $thres$ values that yield models with large number of eliminated
{\em up} states. Since modeling with small errors is fundamental to the
determination of efficient checkpointing intervals, we used $\alpha$
values greater than $\beta$ in our equation for computation of a score
corresponding to a $thres$ value. We performed many experiments with
different values of $\alpha$ and $\beta$ such that $\alpha > \beta$
and chose $\alpha=0.7$ and $\beta=0.3$, since these values resulted in models with
accuracies closer to the original model and with significant number of
eliminated states and hence significant reduction in modeling space
and time complexity. We then find the threshold which has the
maximum score in most of our experiments. Based on these experiments,
we fixed $thres$ as $0.0006$. This threshold of probability resulted
in average number of eliminations of 27-54\% of {\em up} states in our
experiments.

To find the probabilities of transitions in our model, $M^{mall}$, we
construct $N$ birth-death Markov chains, $S^{\tau}$ and $N$
corresponding matrices, $Q^{Up,S}$ and $Q^{Rec,S}$ corresponding to
$N$ different number of active processors used for application execution. Since the
computations of these matrices for a certain number of active
processors are independent of the computations for a
different number of active processors, the construction of the
birth-death Markov chains and the computations in the resulting
matrices for different number of active processors can be parallelized
resulting in reduced execution times of our model. We adopted a
master-worker paradigm where the master program gives the next
available number of processors to a free worker for the calculations
of the corresponding transition probabilities. With these
optimizations, the running time of our model for a given checkpointing
interval is approximately 2-10 minutes. The cost of determining a
checkpointing interval, due to running the model, for an application
execution on a system with a given-failure trace is a one-time cost
for many executions. This is because the selected checkpointing
interval can be used multiple times for the application executions
until the failure rates on the system change significantly.

\section{Rescheduling Policies}
\label{rescheduling}

Our Markov model for malleable applications, $M^{mall}$, is
constructed based on a rescheduling policy that decides the number of
processors for application execution for a given total number of
available processors at a point in the execution. In
this work, we consider three policies for rescheduling. \\
1. {\bf Greedy:} In this policy, when an application recovers after
  failure, it chooses all the available processors for continuing the
  execution. \\
2. {\bf Performance Based (PB):} In this policy, if $a$ is the
  number of processors available for execution, the application
  chooses $n$ processors, $n\leq a$, for which the
  failure-free execution time of the application, $execTime_n$, is
  minimum. \\
3. {\bf Availability Based (AB):} In this policy, the application
  chooses $n$ processors, $n\leq a$, for which the average number of
  failures, $avgFailure_n$, is minimum. To calculate $avgFailure_{n}$
  using a failure trace for a system with a total of $N$ processors,
  $n$ processors are randomly chosen from the $N$ processors in the
  system. The total number of failures for the chosen $n$ processors
  in the trace is calculated as $totalFail$. For calculating
  $totalFail$, a failure is counted if at least one of the $n$
  processors fail at a point of time in the failure trace. $totalFail$
  is then divided by $n$ to obtain $avgTotalFail$. This is repeated
  for 50 different random choices of $n$ processors and the average of
  $avgtotalFail$ values for the 50 random choices is calculated as
  $avgFailure_{n}$.

\section{Experiments and Results}
\label{experiments}

We evaluated our model using three
different applications, three different rescheduling policies and large
number of failure traces. 

\subsection{Failure Data}

For our experiments, we used two kinds of failure traces. One kind of
failure trace corresponds to failure data collected by and available
at Los Alamos National Laboratory (LANL) \cite{lanlfailuredata}. The
data includes the times of failures and repairs of the processors
recorded over a period of 9 years (1996-2005) on 22 different
production high performance computing (HPC) systems at LANL. For our
work, we used two systems, system-1 containing 128 processors and
system-2 containing 512 processors. The second kind of failure trace
corresponds to execution traces of about 740 workstations in the
Condor pool~\cite{condor-web} at University of Wisconsin recorded for
a 18-month period (April 2003~--~October 2004).\footnote{We would like
  to thank Dr. Rich Wolski, UCSB, for providing sanitized Condor
  traces without the host identifiers.} The Condor
project allows execution of {\em guest jobs} on
workstations when they are not used by their owners. When the
workstation owners return, the guest jobs are vacated. For the purpose
of our study, we consider use of a Condor pool for the execution of a
parallel malleable application where the application is a guest job to
the Condor workstations. We thus consider vacation of a guest job in
the Condor trace due to reclaiming of the workstation by its owner as
a failure of the parallel application. The application has to be
checkpointed and continued on a set of free workstations. The
resources in Condor pool are highly volatile with high failure
rates. We consider executing malleable applications on such a volatile
set of disparate resources. The use of such volatile environments for
parallel applications is largely unclear. By conducting our
experiments with the two kinds of failure traces, one corresponding to
dedicated production batch systems and the other corresponding to
highly non-dedicated interactive systems, we attempt to evaluate the
efficiencies of the checkpointing intervals determined by our model
for the different kinds of environments with different failure
rates. We also analyze the variations in the checkpointing intervals
for the two environments.

\subsection{Applications}

We used three different parallel applications. \\
1. ScaLAPACK\cite{scalapack-guide} linear system solver for solving
  over determined real linear systems using QR factorization. The
  specific kernel used was PDGELS. 2-D block cyclic distribution was
  used for the double precision matrix. \\
2. PETSc\cite{petsc-web} Conjugate Gradient (CG) application to solve a system of
  linear equations with a real symmetric positive definite matrix. \\
3. Molecular dynamics simulation (MD) of Lennard-Jones system
  systolic algorithm. N particles are divided evenly among the P
  processes running on the parallel machine. The calculation of forces
  is divided into P stages. The traveling particles are shifted to the
  right neighbor processor in a ring topology. \\
The three applications were executed on a 48-core AMD Opteron
cluster consisting of 12 2-way dual-core AMD Opteron 2218 based 2.64
GHz Sun Fire servers with CentOS 4.3 operating system, 4 GB RAM, 250
GB Hard Drive and connected by Gigabit Ethernet. We assume that the
machines corresponding to the failure traces are similar to the
processors in our cluster. 

The applications were made {\em malleable} by instrumenting them with
function calls to SRS (Stop Restart Software), a user-level
semi-transparent checkpointing library for malleable applications
\cite{vadhiyar-srs-ppl2003}. SRS provides functions for marking data
for checkpointing, reading checkpointed data into variables,
specifying the checkpointing locations and determining if the
application is continued from a previous run. The functions for
marking and reading checkpoint data also allow the users to specify
the data distribution followed for the different variables. By
determining the number of processors and the data distributions, used
for the current and the previous runs, the SRS library automatically
performs the redistributions of data among the current available
processors.
An application is executed for different problem sizes on a certain
number of processors, $a$, for a fixed number of iterations and the
average execution time is calculated. $workinunittime_a$ is then
obtained for the application for $a$ processors by dividing the number
of iterations by the average execution time. The average execution
times for different number of processors were also used to calculate
the rescheduling policy vector, $rp$, for the rescheduling
policies.

To obtain a vector of checkpointing cost, $C$, and a matrix for
recovery cost, $R$, the application instrumented with SRS was executed
with different problem sizes on different number of processors and the
average times for checkpointing or checkpointing overheads were
obtained. These checkpointing overheads were then extrapolated to
larger number of processors using LAB Fit curve fitting
tool\cite{labfit} to form the vector of checkpointing overheads,
$C$. To form the matrix of recovery costs, $R$, the instrumented
application was started on a certain number of processors, $a_1$,
stopped using SRS tools and immediately continued on a different
number of processors, $a_2$. The resulting recovery time between the
stopping and continuing execution on different number of processors is
noted. This was repeated for different problem sizes and the average
of the recovery times was used for the entry, $[a_1,a_2]$ of the $R$
matrix. These costs were then extrapolated for larger number of
processors using LAB Fit.

\begin{figure}
\centering
\includegraphics[width=0.45\textwidth]{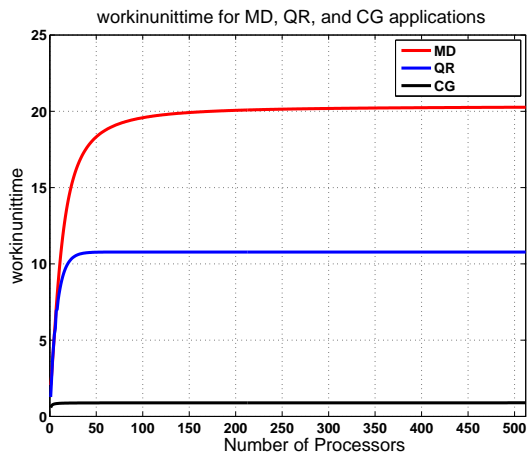}
\caption{$workininuttime$ values for the applications}
\label{wt}
\end{figure}

Figure \ref{wt} shows the number of iterations performed in one second
or the $workininuttime$ values extrapolated to 512 processors for the
three applications. We find that the MD application is highly scalable
and performs more useful work than the other applications. The QR
application with large number of matrix operations is less
scalable than MD while the CG application is the least scalable. Table
\ref{overheads} gives the minimum, average and maximum of the
checkpointing and recovery overheads for the three applications for
different processor configurations. We find that the QR application
has the maximum checkpointing overhead since it checkpoints large
number of large-sized matrices while the MD and CG applications mostly
checkpoint vectors. The MD application has the least checkpointing
overhead since the checkpoints consist of fewer data structures. The
recovery overheads are mostly the same in all applications since
recovery involves simultaneous data redistributions between the
processors.

\begin{table}
\centering
\caption{Checkpointing ($C$) and Recovery ($R$) Overheads}
\begin{tabular}{|p{0.5in}|p{0.3in}|p{0.3in}|p{0.3in}|p{0.3in}|p{0.3in}|p{0.3in}|}
\hline\hline
 & \multicolumn{3}{|c|}{$C$ (secs.)} &
\multicolumn{3}{|c|}{$R$ (secs.)} \\ \cline{2-7}
Application & Min & Avg & Max & Min & Avg & Max \\ \hline\hline
QR & 91.90 & 99.19 & 117.28 & 8.74 & 17.21 & 32.97 \\ \hline
CG & 8.96 & 9.55 & 9.75 & 8.89 & 12.56 & 15.12 \\ \hline
MD & 1.35 & 1.84 & 2.70 & 8.27 & 14.12 & 17.05 \\ \hline\hline
\end{tabular}
\label{overheads}
\end{table}

\subsection{Evaluation}

For an application with a given $workinunittime$, $C$, $R$, and $rp$,
our models were invoked for different {\em execution segments} for the
failure traces of a system. A given execution segment for a failure
trace on a system of $N$ processors, corresponds to
a random start time, $start$, and random duration, $dur$, for
application execution. Based on $start$, $\lambda$ and
$\theta$ are calculated for the execution segment using the history of
failures that occurred on the processors before $start$. Our model,
$M^{mall}$, is invoked with the parameters, $workinunittime$, $C$,
$R$, $rp$, $N$, $\lambda$ and $\theta$ for a given application, and
execution segment for a failure trace of a system, with different
values of $I$.

We have developed a simulator to assess the quality of the checkpointing
intervals determined by our model for an application execution on an
execution segment. Our simulator uses the same inputs
to our model along with duration, $dur$, for simulating the
application execution that started at $start$ point in the failure
trace and executed for $dur$ seconds. At $start$, the simulator
considers the number of available processors and based on the
rescheduling policy, $rp$, chooses a set of available processors,
$a_1$, for application simulation. The simulator then simulates
application execution by advancing the time and accumulating
checkpointing intervals, $I$ to the useful computation time,
$u_{a_1}$. After every $I$ seconds, the simulator simulates
application checkpointing by advancing $C_{a_1}$ seconds in time. This
is repeated until a failure occurs in one of the $a_1$ processors. At
this point, the simulator calculates the amount of useful work spent
on $a_1$ processors, $uw_{a_1}$ as $uw_{a_1} =
workinunittime_{a_1}\times u_{a_1}$. It accumulates $uw_{a_1}$ to
total amount of useful work, $UW$. After failure, the simulator once
again considers the number of available processors. If none of the
processors are available, the simulator adds the time corresponding to
waiting for one of the processors to be repaired for application
continuation. At that point in time, the simulator considers the total
number of available processors, and based on the rescheduling policy,
$rp$, chooses another set of available processors, $a_2$, for
application execution. Based on the previous and current set of
processors for application execution, $a_1$ and $a_2$, respectively,
the simulator advances the time by $R_{a_1,a_2}$ seconds for
recovery. After recovery, the simulator once again advances time by
accumulating checkpointing intervals, $I$, until one of the processors
in $a_2$ fails at which point it calculates the useful work spent in
$a_2$ processors, $uw_{a_2}$, and adds this to total useful time,
$UW$. This process continues until the simulator reaches $start+dur$
seconds of time, at which point it outputs the total useful work
performed by the application as $UW$.

For a given application execution in an execution segment of a failure
trace, we calculate {\em model efficiency}. To find model efficiency,
we find the checkpointing interval, $I_{model}$, corresponding to
large $UWT_{model}$ values produced by the model. We then find the
useful work output by the simulator,  $UW_{I_{model}}$, for the
execution segment corresponding to $I_{model}$. We also find the
highest useful work, $UW_{highest}$, output by the simulator for the
execution segment when executed with different values of checkpointing
interval. We denote the checkpointing interval corresponding to
$UW_{highest}$ as $I_{sim}$. We find the percentage difference, $pd$,
between $UW_{highest}$ and $UW_{I_{model}}$ to give the percentage of
work lost due to executing the application with the interval
determined by our model. The percentage difference, $pd$, represents
the {\em model inefficiency}, while $(100-pd)$ represents the {\em
  model efficiency}.

Due to modeling errors, instead of choosing the interval corresponding
to the highest $UWT_{model}$ value, we choose the intervals
corresponding to $UWT_{model}$ values that are within 8\% of the
highest $UWT_{model}$. We then calculate the average of these
intervals as $I_{model}$. For exploring different checkpoint intervals
to determine $I_{model}$, we use a minimum checkpoint interval,
$I_{min}$. The checkpointing intervals are doubled starting from
$I_{min}$ until the $UWT_{model}$ for the current checkpoint interval
is less than the value for the previous interval. We then perform
binary-search within the intervals corresponding to the top three
$UWT_{model}$ values to explore more checkpointing intervals
corresponding to high $UWT_{model}$ values. For this work, we use 5
minutes for $I_{min}$.

\subsection{Results}

Table \ref{different-systems} shows the model efficiency for different
number of processors on different systems for QR application with greedy rescheduling
policy. In all cases, the efficiency of our model in terms of the
amount of useful work performed by the application in the presence of
failures, using the intervals determined by our model, as shown
in column 5, was greater than 80\%. Thus, the intervals determined
using our models are highly efficient for executing malleable parallel
applications on systems with failures. We also find that the
checkpointing intervals determined by our model increases with
decrease in failure rates of systems indicating that the model
presents practically relevant checkpointing intervals. The average
$UWT$ values, determined using the simulator, corresponding to
$I_{model}$ and $I_{sim}$ are comparable and follow similar trends for
the different systems. This shows that the checkpointing intervals
determined by our models are highly competent with the best
checkpointing intervals. The intervals determined by our model are
smaller for the Condor systems than for the batch systems, with the
interval approximately equal to 35 minutes when considering execution
of malleable parallel applications on a Condor pool of 256
processors. This is due to the highly volatile non-dedicated Condor
environment as indicated by the higher failure rates or $\lambda$s for
the Condor systems. However, we find that our model
efficiencies for the Condor systems are equivalent to the
efficiencies for the batch systems, implying that our model can
determine efficient checkpointing intervals for both dedicated batch
systems and non-dedicated interactive systems.

\begin{table*}
\centering
\caption{Model Efficiencies for Different Systems (QR application,
  greedy rescheduling policy)}
\begin{tabular}{|p{0.2in}|p{0.5in}|p{0.8in}|p{0.9in}|p{0.65in}|p{0.45in}|p{0.45in}|p{0.45in}|}
\hline\hline
Procs. & System & Average $\lambda$ & Average $\theta$ & Average Model Efficiency
\% & Average $I_{model}$ (hours) & Average $UWT$ for
$I_{model}$ & Average $UWT$ for $I_{sim}$
\\ \hline\hline
64  &  system-1 & 1/(6.42 days)     & 1/(47.13 min.) & 80.17 & 2.81    & 8.27 & 9.45
\\ \hline
128 & system-1 & 1/(104.61 days) &  1/ (56.03 min.) & 90.37 & 17.78  & 9.57 & 10.46
\\ \hline
256 & system-2 & 1/(81.82 days)   &  1/(168.48 min.) & 86. 14 & 5.32    & 8.67 & 9.85
\\ \hline
512 & system-2 & 1/(68.36 days)   &  1/(115.43 min.) & 95.74 & 3.68    & 9.76 & 10.17
\\ \hline
64 &   condor    & 1/(6.32 days)      & 1/(52.377 min.) & 82.33 & 2.75    & 8.44 & 9.52
\\ \hline
128 & condor    & 1/(6.36 days)      &  1/(54.848 min.) & 87.19 & 1.53   & 8.26 & 9.08
\\ \hline
256 & condor    & 1/(5.19 days)      & 1/(125.23 min.)  & 93.38 & 0.67   & 7.89 & 8.32
\\ \hline\hline
\end{tabular}
\label{different-systems}
\end{table*}

We also compared the model efficiencies for the three applications on
128 processors of system-1 with greedy rescheduling strategy. Table
\ref{app-results} shows the efficiencies for the applications. The table shows that the
checkpointing intervals by our models are more than 90\% efficient in
terms of the amount of useful work performed by the applications. This
shows that our modeling strategy is applicable to different
applications. We find that the checkpointing intervals determined by
our model, $Imodel$, are largest for the QR application. As shown in
Table \ref{overheads}, QR application has high checkpointing and
recovery overheads. Hence our modeling strategy tries to maximize the
amount of useful work for the QR application by selecting larger
checkpointing intervals resulting in smaller non-useful or down times
for the application. We find that the $UWT$ values corresponding to
$I_{model}$ and $I_{sim}$ are comparable. For all the three
applications, the $UWT$ values calculated for application executions in
the presence of failures are within 4-11\% of the corresponding
failure-free maximum $workinunittime$ values shown in Figure
\ref{wt}. This shows that by adopting malleability, and choosing
efficient checkpointing intervals by our models, applications can
execute with nearly failure-free high performance even in the presence
of failures.
\begin{table}
\centering
\caption{Model Efficiencies for the 3 Applications (system-1, 128
  processors, greedy rescheduling policy)}
\begin{tabular}{|p{0.6in}|p{0.5in}|p{0.4in}|p{0.4in}|p{0.4in}|}
\hline\hline
Application & Average Model Efficiency \% & Average $I_{model}$ (hours) &
Average $UWT$ for $I_{model}$ & Average $UWT$ for $I_{sim}$
\\ \hline\hline
QR & 90.37 & 17.78 & 9.57 & 10.47 \\ \hline
CG & 95.66 & 7.59 & 0.85 & 0.88 \\ \hline
MD & 90.06 & 13.7 & 17.96 & 19.72 \\ \hline\hline
\end{tabular}
\label{app-results}
\end{table}

\begin{figure}
\centering
\includegraphics[width=0.45\textwidth]{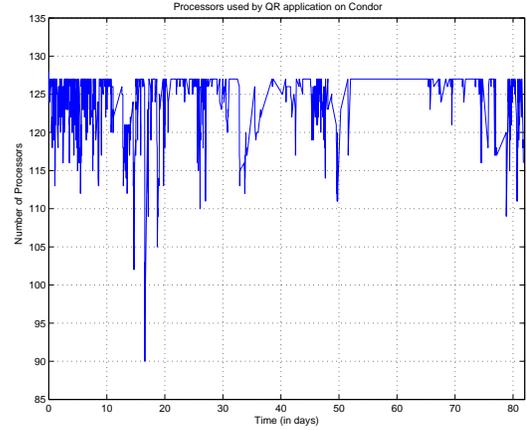}
\caption{Simulation of QR application execution on Condor system for
  80 days (128 processors, checkpointing interval = $I_{model}$ = 1.53 hours,
  $C=R=$ 20 minutes)}
\label{condor-sample}
\end{figure}

We also investigated the usefulness of non-dedicated highly-volatile
Condor environments for the execution of malleable parallel
applications. The work by Plank and
Thomason\cite{plank-processorallocation-jpdc01} showed that Condor
systems, due to high failure rates and checkpointing overheads, are
not suitable for execution of moldable applications since a fixed
number of processors are generally not available for the entire
duration of application execution on such systems. Accordingly, it was
shown that execution on only one processor in Condor environment
provided the least runtimes for the applications. We explored the use
of such systems for malleable applications. Figure \ref{condor-sample}
shows simulation of a sample execution of QR application on 128
processors of the Condor system for a duration of 80 days with the
greedy rescheduling policy. $I_{model}$ of 1.53 hours, the
checkpointing interval determined by our model, was used for the
execution. We also used $C=R=20$ minutes, the worst-case checkpointing
and recovery overheads on shared Condor systems and networks.

As Figure \ref{condor-sample} shows, different number of processors
are used for execution at different times. More than 100 processors
are used in most cases since the greedy rescheduling strategy chooses
the maximum number of processors available for execution. The $UWT$
value for this execution is 7.29. This $UWT$ value, obtained in the
presence of failures on the Condor systems, is nearly 70\% of the
corresponding failure-free maximum $workinunittime$ value for QR
application shown in Figure \ref{wt}. Thus, while highly volatile
environments like Condor are not suitable for executions of moldable
applications,  they can be used to provide high efficiency for
malleable applications due to the flexibility in terms of the number
of processors used for execution and the efficient checkpointing intervals
determined by our models.

Table \ref{different-rp} shows the model efficiency and the amount of
useful work corresponding to the intervals determined by our models
for the three rescheduling policies for QR application on 128
processors of system-1. We find that in all the rescheduling policies,
the efficiency was greater than 80\%. We also find that the AB
rescheduling policy yields the maximum work for the application. This
is because the policy attempts to execute the application on small
number of processors where the mean time to failures is low. Hence
larger checkpointing intervals are chosen for application execution,
as shown in the table, leading to less checkpointing overheads and
more useful work. The greedy strategy performs the least useful work
since it always executes the application on maximum number of
available processors where the mean time to failures is high leading
to high checkpointing and recovery overheads. The PB policy considers
number of processors for which failure-free running time is
minimum. Since the QR application is highly scalable, the PB
rescheduling policy attempts to execute on large number of processors
where the failure rates are high. Hence, the checkpointing intervals
and the resulting amount of useful work for the PB policy are
comparable to those for the greedy policy. Thus, we find that for
large number of systems with failures, executing on smaller number of
systems with less failure rates (AB) leads to more useful work by the
application than executing the application on the number of processors
corresponding to maximum performance (PB).

\begin{table}
\centering
\caption{Model Efficiencies for Different Rescheduling Policies (QR
  application, system-1, 128 processors)}
\begin{tabular}{|p{0.6in}|p{0.7in}|p{0.55in}|p{0.55in}|}
\hline\hline
Resched. Policy & Average Model Efficiency
\% & Average $I_{model}$ (hours) & Average $UW_{I_{model}}$  $(\times
10^6)$ 
\\ \hline\hline
Greedy & 90.3   & 17.41 & 108.27 \\ \hline
PB        & 90.0 & 17.44 & 110.20 \\ \hline
AB        & 83.1 & 88.42 & 133.15 \\ \hline
\hline
\end{tabular}
\label{different-rp}
\end{table}

\begin{figure}
\centering
\subfigure[Different Failure  Rates (256 processors)]{
  \includegraphics[width=0.45\textwidth]{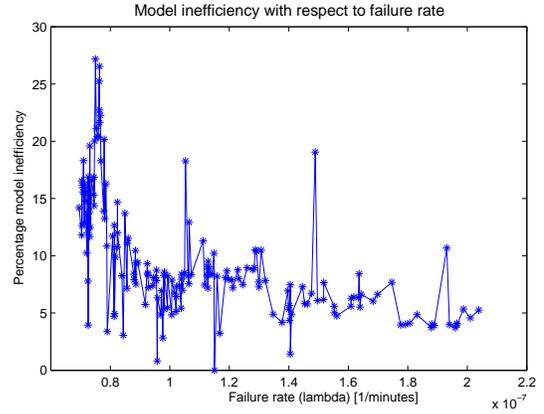}
  \label{rate}
}
\hfil
\subfigure[Different Durations (128 processors)]{
  \includegraphics[width=0.45\textwidth]{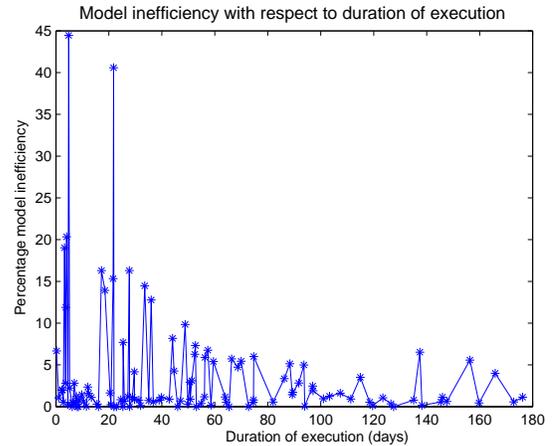}
  \label{dur}
}
\caption{Model Efficiency for Different Failure Rates and Durations (QR application,
  condor trace, greedy rescheduling policy)}
\label{rate-dur}
\end{figure}

Figure \ref{rate} shows the model inefficiencies with increasing
failure rates for QR application on a condor trace with 256 processors
and for greedy rescheduling policy. The figure shows that the model
inefficiencies decrease or model efficiencies increase with increasing
failure rates. Thus our model is effectively able to predict
application behavior on systems with frequent failures than for
systems with sporadic failures. This is because the history of failure
rates on systems with sporadic failures cannot be effectively used to
predict the future failures on those systems. We also compared the
model efficiencies for varying durations of application
execution. Figure \ref{dur} shows the results for QR application
execution with condor traces on 128 processors and for greedy
rescheduling policy. The results show that our model inefficiency
decreases or model efficiency improves with increasing durations. This
is because the long-running properties of our Markov model are
especially suited for long-running parallel applications than for
short applications.

\subsection{Summary}

In addition to showing validation and efficiency results, our
experiments have also shown some interesting and new observations. We
have shown that with the help of malleability and our checkpointing
intervals, applications can execute with near failure-free performance
in the presence of failures. We have also shown that malleability and
our checkpointing intervals encourage executions on volatile
environments like Condor, the environments considered to be not
suitable for parallel applications in the earlier efforts.

\section{Related Work}
\label{related}

There is a vast amount of literature on checkpointing interval
selction\cite{elnozahy-survey-cmureport96}. In this section, we focus
on some of the highly relevant efforts. The work by
Daly\cite{daly-higherorder-fgcs06} develops a first-order model to
determinine optimum checkpointing interval. The work then develops a
higher-order model for improving accuracy for small talues of
MTBF. The work assumes Poisson failure rate and considers single
processor failures.

The work by Nurmi et al.\cite{nurmi-modelbased-ucsbreport04}
determines checkpointing intervals that maximize efficiency of an
application when executed on volatile resource-harvesting systems such
as Condor\cite{condor-web} and SETI@Home\cite{setiathome}. They use
three different distributions including exponential, Weibull and
hyper-exponential for fitting hostorical data on machine
availabilty. They then use the future lifetime distribution along with
the checkpoint parameters including checkpointing intervals in a
three-state Markov model for application execution and determine the
interval that minimize execution time. Their Markov model is based on
the work by by Vaidya\cite{vaidya-impact-itc97}. Their work is
intended for sequential processes executing in the Condor
environment. They show that the type of failure distribution does not
affect the application execution performance.

The work by Ren et al.\cite{ren-failureaware-hpdc07} considers proving
fault tolerance for {\em guest jobs} executing on resources provided
voluntarily in fine grained cycle sharing (FGCS) systems. In their
work, they calculate checkpointing interval using a low overhead
one-step look ahead heuristic. In this heuristic, they divide
execution time into steps and compare the costs of checkpointing at a
step and the subsequent step using the probability distribution
function for failures. Their model does not assume a specific
distribution and is intended for sequential guest jobs.

Plank and Elwasif\cite{plank-experimentalassessment-ftcs98} study the
implications of theoretical results related to optimal checkppointing
intervals on actual performance of application executions in the
presence of failures. They perform the study using simulations of long
running applications with failure traces obtained on three parallel
systems. One of the primary results of their work is that the
exponential distribution of machine availability, although inaccurate,
can be used for practical purposes to determine checkpointing
intervals for parallel applications. 

To our knowledge, the model by Plank and
Thomason\cite{plank-processorallocation-jpdc01} is the most
comprehensive model for determining checkpointing intervals for
parallel applications. Their work tries to determine the number of
processors and checkpointing interval for executing moldable parallel
applications on a parallel system given a failure trace on the
system. The model assumes exponential distribution for inter-arrival
times of failures. They use the concept of spare processors for
replacing the systems that failed during application execution. They
show by means of simulations that checkpointing intervals determined
by their model lead to reduced execution times in the presence of
failures. Their work was intended for typical parallel checkpointing
systems that do not allow the number of processors to change during
application execution. With increasing prevelance of checkpointing
systems for malleable parallel applications, we base our work on their
model and significantly modify different aspects of their model to
determine checkpointing intervals for such systems.

\section{Conclusions}
\label{conclusions}

The work described in this paper presents the first effort, to our
knowledge, for selecting checkpointing intervals for efficient
execution of malleable parallel applications in the presence of failures. The work
is based on a Markov model for malleable applications that includes
states for execution on different number of processors during
application execution. We have also defined a new metric for
evaluation of such models. The states of our model are based on
rescheduling policies. By conducting large number of simulations with
failure traces obtained for real high performance systems, we showed
that the checkpointing intervals determined by our model lead to
efficient executions of malleable parallel applications in the
presence of failures.

\section{Future Work}
\label{futurework}

We plan to augment our model with different kinds of failure
distributions. We also plan to experiment with different
redistribution policies. The checkpoint intervals from our model will
be integrated with a real checkpointing system that provides
malleability and fault tolerance. Finally, we plan to extend our model
for determining checkpointing intervals for executions on
multi-cluster grids and heterogeneous systems.

\bibliographystyle{IEEEtran}
\bibliography{ckptinterval}

\end{document}